\documentclass[a4paper,12pt]{article}
\usepackage{amsmath}
\usepackage{epsfig}
\newcommand{\Dzero}{D\O{} }
\newcommand{\etal}{{\em et al.}}

\newcommand{\slE}{\mbox{/}\!\!\!\!E}
\newcommand{\GeV}{\mbox{ GeV}}

\def\slashchar#1{\setbox0=\hbox{$#1$}           % set a box for #1
   \dimen0=\wd0                                 % and get its size
   \setbox1=\hbox{/} \dimen1=\wd1               % get size of /
   \ifdim\dimen0>\dimen1                        % #1 is bigger
      \rlap{\hbox to \dimen0{\hfil/\hfil}}      % so center / in box
      #1                                        % and print #1
   \else                                        % / is bigger
      \rlap{\hbox to \dimen1{\hfil$#1$\hfil}}   % so center #1
      /                                         % and print /
   \fi}

\newcommand{\slptwo}{\slashchar{{\bf p}}}
\newcommand{\slp}{\slashchar{p}}
\newcommand{\ptwo}{{\bf p}}

\begin{document}
\begin{titlepage}
\vspace*{-1cm}
\begin{flushright}
Cavendish--HEP--99/07\\
June 1999\\
\end{flushright}                                

\vskip 1.cm
\begin{center}                                                             

{\Large\bf
Measuring masses of semi-invisibly decaying\\
particles pair produced at hadron colliders 
}
\vskip 1.3cm
{\large C.~G.~Lester\footnote{email: lester@hep.phy.cam.ac.uk} 
and  D.~J.~Summers\footnote{email: summers@hep.phy.cam.ac.uk} }
\vskip .2cm
{\it High Energy Physics Group, Cavendish Laboratory,\\
         Madingley Road, Cambridge CB3 0HE, England}
\vskip 2.3cm   

\end{center}      

\begin{abstract}
We introduce a variable useful for measuring masses of particles pair
produced at hadron colliders, where each particle decays to one
particle that is directly observable and another particle whose
existence can only be inferred from missing transverse momenta. This
variable is closely related to the transverse mass variable commonly
used for measuring the $W$ mass at hadron colliders, and like the
transverse mass our variable extracts masses in a reasonably model
independent way. Without considering either backgrounds or measurement
errors we consider how our variable would perform measuring the mass
of selectrons in a mSUGRA SUSY model at the LHC.
\end{abstract}                                                                

\vfill
\end{titlepage}
\newpage                                                                     

Before 1983, $e^+ e^-$ colliders dominated the search for new
particles, their clean final states making identification of new
particles far easier than at hadron machines.
%Prior to the experimental discovery of the $W$ \cite{ua1w,ua2w} and
%$Z$ \cite{ua1z,ua2z} bosons in 1983
%all discoveries of fundamental particles had been made at
%$e^+e^-$ and  $ep$ machines with clean initial state
%particles. 
However in 1983 this changed
with the observation of the $W$ and $Z$ bosons by the UA1\cite{ua1w,ua1z} and
UA2\cite{ua2w,ua2z} collaborations at the CERN $p\bar p$
collider. Despite the more difficult conditions where
the initial state protons and anti-protons gave rise to an underlying
event which complicates measurement of the final state, the higher
energies available to a $p\bar p$ machine meant they discovered the
$W$ and $Z$ bosons before other machines had enough energy to produce
them. This trend continued in 1994/5 when CDF and
\Dzero discovered the $t$ quark at the Tevatron $p\bar p$ collider
\cite{top}.  To date the only direct experimental evidence that we
have for the $t$ quark is from the Tevatron.

Clearly, the greater energies available to $p\bar p$ machines gives
them an advantage in discovering fundamental particles, and we can
hope that this trend will be continued at the LHC. Upon discovery of a
new particle the first question that is asked is what is the mass of
this particle. There are two fundamentally different ways of
measuring the mass of a particle. Usually one tries to
extract the mass of the particle directly from the observed events. For example
UA1 and UA2 detected the $Z$ boson in its decay to lepton pairs $e^+
e^-$ or $\mu^+ \mu^-$. For each event that contains a pair of leptons
one can construct the mass of the particle that produced the leptons
as,
\begin{equation}
m^2=(p_{l^+}+p_{l^-})^2
\end{equation}
and on an event by event basis obtain a direct estimate the mass of the
particle. Alternatively one
could make an indirect model dependent measure of the $Z$ mass by
measuring the ``$Z$ \nolinebreak event'' cross-section, and using a model that
predicts the $Z$ cross-section in terms of the $Z$ mass, perform a
fit to the measured cross-section. For measuring the  $Z$ mass the direct
measurement is superior, if only for its lack of model
dependence, however when measuring the mass of the $W$ boson the case
is less clear. At the CERN $p\bar p$ collider the $W$ boson was
detected in its decay to a charged lepton and a neutrino, however the
neutrino goes unobserved and so only gives rise to missing
momentum. This means that one can't directly observe the $W$ mass from
the lepton and neutrino momenta. UA1 \cite{ua1w} formed the transverse
mass variable, 
\begin{equation}
m_T^2=2(E_T^e \slE_T - {\bf p}_T^e \cdot \slptwo_T) ,
\label{eq:mt}
\end{equation}
where $\slE_T^2 \equiv \slptwo_T^2$. This variable has the property that,
\begin{equation}
m_T^2 \le m_W^2
\end{equation}
with equality possible for events where the lepton and neutrino are
produced with the same rapidity. Thus although one can't obtain the
$W$ mass from 
a single event, one can obtain a lower limit on the $W$ mass. In
addition if one obtains the lower mass bound from many events, this
can approach the $W$ mass, and so one can extract the $W$ mass in a
model independent way. UA1 \cite{ua1w} along with UA2 \cite{ua2w} also
performed a model depenendent fit to the $p_T$ spectrum for the
lepton, extracting the $W$ mass from that model.

% In practice though the distinction between these two
% variables, $M_T$ and $p_T(l)$ for measuring the $W$ mass is not as
% clear as we made out above, precisely what what fraction of events
% have an $M_T$ close to $m_W$ is still dependent on
% the physics process which produces the $W$ boson. 
% This gives the mass extracted
% from the transverse mass some small model dependence. 

In practice, though, measuring the $W$ mass with $m_T$ does 
have some small model dependence, 
as the precise fraction of events which occur with $m_T$ close to $m_W$
is still dependent on the physics processes which produce the $W$
boson, and how the $W$ boson decays.
In addition, the missing transverse momentum is  poorly measured
experimentally compared  with $\ptwo_T(l)$, so the
theoretical model dependence of the measurement of $m_W$ from the
$\ptwo_T(l)$ spectrum is balanced by the experimental error on extracting
$m_W$ from the edge of the $m_T$ spectrum.

\begin{figure}
\begin{center}
\psfig{figure=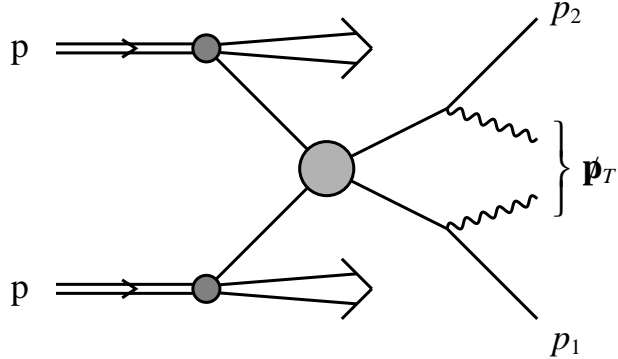,height=5cm}
\end{center}
\caption{Diagram of the generic process that we consider. A hadronic
	collision that leads to a pair of particles being produced,
	which each decay into one particle that is observed with
	momenta $p_1$ and $p_2$ respectively; and one particle (shown
	as a wavy lines) that is not directly detected, and whose
	presence can only be inferred from the missing transverse
	momentum, $\slptwo_T$.
  \label{fig:generic}}
\end{figure}

In this paper we wish to introduce a variable which measures particle masses,
which like transverse mass has little dependence on exactly how such
massive particles are produced. The variable is used for the generic process
shown in figure \ref{fig:generic}, where a hadronic collision pair produces a
massive particle whose dominant decay is into one observed and one unobserved
particle. This unobserved particle can only be detected from the missing
momentum that it carries away, and that the massive particle is pair produced
means that we can only measure the missing momentum of the pair of invisible
particles. Although this may sound like an unusual process to look for new
particles, it naturally occurs in any theory where there is an
(approximately) conserved charge, and the lightest particle with that charge
is only weakly interacting. Two examples of where such a situation can occur
are SUSY models and a 4th lepton generation. In R-parity conserving SUSY
models, sparticles are pair produced, and cascade decay to the lightest
sparticle, which must be stable and is expected to not be directly
detectable. Slepton production and decay can often follow this route:
\begin{equation}
pp \to X + \tilde{l}^+_R \tilde{l}^-_R 
\to X + l^+ l^- \tilde{\chi}^0_1 \tilde{\chi}^0_1 .
\label{eq:susy}
\end{equation}
In such an event the pair of lightest SUSY particles, $\tilde{\chi}$, go
unobserved and only leave their signature as missing transverse
momentum.

For a 4th generation lepton the charged lepton would be pair produced
in a Drell-Yan type process, decaying to a neutrino and a
$W$ boson, 
\begin{equation}
pp \to X + l_4^+ l_4^- \to X + \bar\nu_{l_4} W^+ \nu_{l_4} W^-
\end{equation}
the (probably massive) $\nu_{l_4}$ going unobserved, while the $W$
bosons could be detected in their decays to either $l\nu$ or to jets.

We now look specifically at the process given in equation
\ref{eq:susy}, although the variable which we
define would work identically in any process where a particle is pair
produced and decays to one visible and one invisible particle.

The variable that we wish to introduce is closely related to $m_T$, 
however the standard definition of $m_T$, given in equation
\ref{eq:mt}, assumes 
that the unobserved particle is massless, so we return to the
derivation of this variable. For the decay,
\begin{equation}
\tilde{l}\to l \tilde{\chi}
\end{equation}
for arbitrary momenta we can write,
\begin{equation}
m_{\tilde{l}}^2 = m_l^2 + m_{\tilde{\chi}}^2 + 
2 ( E_{Tl} E_{T\tilde{\chi}} \cosh ( \Delta\eta ) 
      - {\bf p}_{Tl}\cdot{\bf p}_{T\tilde{\chi}})
\end{equation}
where $E_T=\sqrt{{\bf p}_T^2 + m^2}$  and
$\Delta\eta$ is the difference in rapidity,
$\eta=\frac{1}{2}\ln[(E+p_z)/(E-p_z)]$,  between between the $l$ and
$\tilde{\chi}$.

Now as $\cosh\eta \ge 1$ we have,
\begin{equation}
m_{\tilde{l}}^2 \ge m_T^2(\ptwo_{Tl},\ptwo_{T\tilde{\chi}}) 
\equiv m_l^2 + m_{\tilde{\chi}}^2 + 
2 ( E_{Tl} E_{T\tilde{\chi}} - \ptwo_{Tl}\cdot\ptwo_{T\tilde{\chi}}).
\label{eq:mtd}
\end{equation}
This gives a version of transverse mass valid for arbitrary masses,
 with equality when the
$l$ and $\tilde{\chi}$ are produced with the same rapidity. 
Notice that $E_{Tl}$ and $E_{T\tilde{\chi}}$ depend on 
$m_l^2$ and $m_{\tilde{\chi}}^2$ respectively.

The transverse mass can't be formed directly from the process in equation
(\ref{eq:susy}), as both the neutralinos give rise to missing momentum,
however we can experimentally measure the sum of their transverse
momenta as the missing transverse momenta in the event,
\begin{equation}
\slptwo_T = \ptwo_{T\tilde{\chi}_a} + \ptwo_{T\tilde{\chi}_b}.
\label{eq:ptmsum}
\end{equation}
If $\ptwo_{T\tilde{\chi}_a}$ and $\ptwo_{T\tilde{\chi}_b}$ were obtainable,
then one could form two transverse masses, and using the relationship
(\ref{eq:mtd}) obtain,
\begin{equation}
m_{\tilde{l}}^2 \ge {\rm max} \{ m_T^2(\ptwo_{Tl^-},\ptwo_{T\tilde{\chi}_a}) ,
			m_T^2(\ptwo_{Tl^+},\ptwo_{T\tilde{\chi}_b}) \}
\end{equation}
However, not knowing the form of the splitting (\ref{eq:ptmsum}), the best we
can say is that:
\begin{equation}
m_{\tilde{l}}^2 \ge M_{T2}^2 \equiv \underset{
% {\rm splittings } 
%	\slptwo_1 +\slptwo_2=\slptwo_T }{\min} 
	/\!\!\!\!{\bf p}_1 +/\!\!\!\!{\bf p}_2=/\!\!\!\!{\bf p}_T }{\min} 
	\Bigl[
	\max{   \{ m_T^2(\ptwo_{Tl^-}, \slptwo_1 ) ,
		      m_T^2(\ptwo_{Tl^+}, \slptwo_2 ) \} }
	\Bigr]
\label{eq:mt2}
\end{equation}
With the minimization over all possible 2-momenta, $\slptwo_{1,2}$,
such that their sum gives the observed missing transverse momentum,
$\slptwo_T$.
This is the variable, called $M_{T2}$, that we wish to introduce.
This bound we can obtain directly from experimentally measured
parameters. Although not totally transparent, for particular momenta,
$M_{T2}$ can be equal to $m_{\tilde{l}}$; the requirement being that
for both slepton decays the lepton and neutralino are
produced at the same rapidity (although the sleptons themselves can be
at differing rapidities), and in addition,
\begin{equation}
\Bigl(\frac{\ptwo_{Tl^-}}{E_{Tl^-}} -  
      \frac{\ptwo_{T\tilde{\chi}_a}}{E_{T\tilde{\chi}_a}} \Bigr)
\propto
\Bigl(\frac{\ptwo_{Tl^+}}{E_{Tl^+}} -  
      \frac{\ptwo_{T\tilde{\chi}_b}}{E_{T\tilde{\chi}_b}} \Bigr).
\end{equation}
%Since the splitting of $\slptwo_T$ into $\slptwo_1$ and
%$\slptwo_2$ typically occurs for momenta that the two neutralinos
%could have had, events can have $M_{T2}$
%arbitrarily close to $m_{\tilde{l}}$. 

We have not managed to derive a general analytic expression for the
minimization over splittings of $\slptwo_T$; largely because an
experimental measurement of $\slptwo_T$ only measures the missing
transverse momentum, and neither the missing energy nor the missing
longitudinal momenta. This means that $\slptwo_T$ is not a 4 vector,
which means that $M_{T2}$ can not be calculated in a manifestly
Lorentz invariant manner.
The complication that this introduces to the 
minimization is enough to making an analytic solution non
trivial. However if we take one of the parameters that we minimize
over to be $m^2=\slp_T^2$ ({\em i.e.} that
$\slE_T=\sqrt{m^2+\slptwo_T^2}$), and in addition note that longitudinal
momenta play no part in the definition of $M_{T2}$,
then the remaining minimization becomes Lorentz invariant, and this means
that it can be solved analytically. This leaves one minimization over
$m^2$ that we performed numerically. The form of this
minimization is not particularly illuminating, and so we do not give
it here; however a computer code for evaluating $M_{T2}^2$ is
available from the authors. 

Of course, the variable $M_{T2}$ is only good at extracting particle
masses from processes having many events close to the maximum allowed
value, and this depends on the physics of the process being
measured. Hence, a priori, we can not say that $M_{T2}$ is useful for
measuring particle masses in all processes. However we expect typical
physics processes to have reasonable numbers of events close to the
maximum value allowed for $M_{T2}$ and hence expect it to be a useful
variable.  To illustrate the variable in use, we consider a model
which allows the process shown in equation (\ref{eq:susy}), at the
expected LHC centre of mass energy
$\sqrt{s}=14\mbox{TeV}$. Our SUSY model is the fifth minimal supergravity model (mSUGRA)  
\cite{susymodel} point selected by the LHC Committee in 1996
for detailed study by the ATLAS and CMS
collaborations \cite{www}.  This model is characterized by,
\begin{equation}
\{\tan\beta=2.1, \ m_{1/2}=300\GeV, \ m_0=100\GeV, \ A_0=300\GeV, \ \mu > 0 \}.
\end{equation} 
We generate events
using the SUSY version of the Herwig generator \cite{herwigino}. For our
purposes the only important features of the model that we use to
produce our events are,
\begin{equation}
m_{\tilde{l}_R}=157.1\ {\rm GeV} \qquad m_{\tilde{\chi}^0_1}=121.5\ {\rm GeV}.
\end{equation}
The angular distributions for the sleptons are given by the Drell-Yan
process that produces spin-0 particles, while the decay of the
sleptons is isotropic due again to their being spinless. Herwig serves
to build up a reasonably realistic underlying event, {\em i.e.} 
that $\slptwo_T \neq {\bf p}_{Tl^+} +  {\bf p}_{Tl^-}$.
Defining the missing momenta as,
\begin{equation}
\slptwo_T={\bf p}_{T\tilde{\chi}_a} +  {\bf p}_{T\tilde{\chi}_b},
\end{equation}
we generate 1105 events, which corresponds to an integrated
luminosity of about $30\mbox{ fb}^{-1}$, which the LHC should collect in
approximately one year of low luminosity running.

\begin{figure}
\begin{center}
\psfig{figure=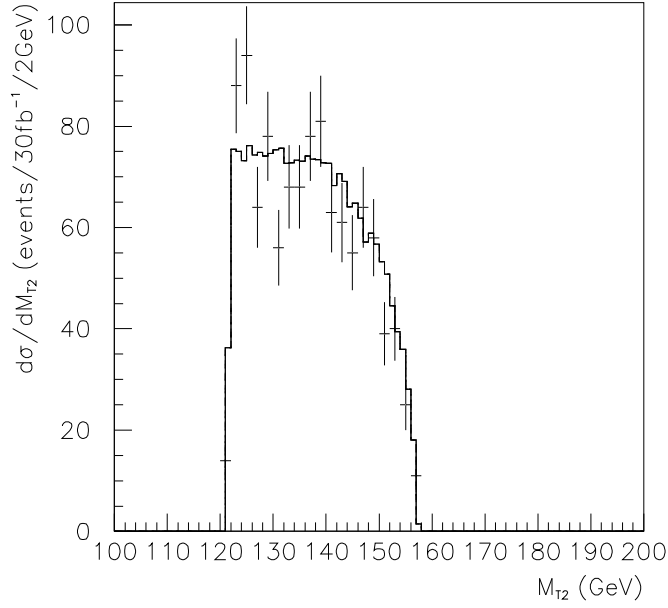,height=8cm}
\end{center}
\caption{$M_{T2}$ distribution for the process 
	$pp \to X + \tilde{l}^+_R \tilde{l}^-_R 
	\to X + l^+ l^- \tilde{\chi}^0_1 \tilde{\chi}^0_1$
	at the LHC. With $m_{\tilde{l}}=157.1\mbox{ GeV}$ and 
	$m_{\tilde{\chi}}=121.5\mbox{ GeV}$, assuming the actual value for
	$m_{\tilde{\chi}}$ when calculating $M_{T2}$. The data with
	error bars are 1105 events, that the LHC would collect in
	approximately 1 year of running at low luminosity, {\em i.e.}
	${\cal L}\simeq 30\mbox{ fb}^{-1}$. The histogram represents 
	${\cal L}\simeq 500\mbox{ fb}^{-1}$ to show the shape of the
	distribution that would be obtained with huge statistics, with
	the normalization modified to be the same as 
	${\cal L}\simeq 30\mbox{ fb}^{-1}$.
  \label{fig:mt2}}
\end{figure}

\begin{figure}
\begin{center}
\psfig{figure=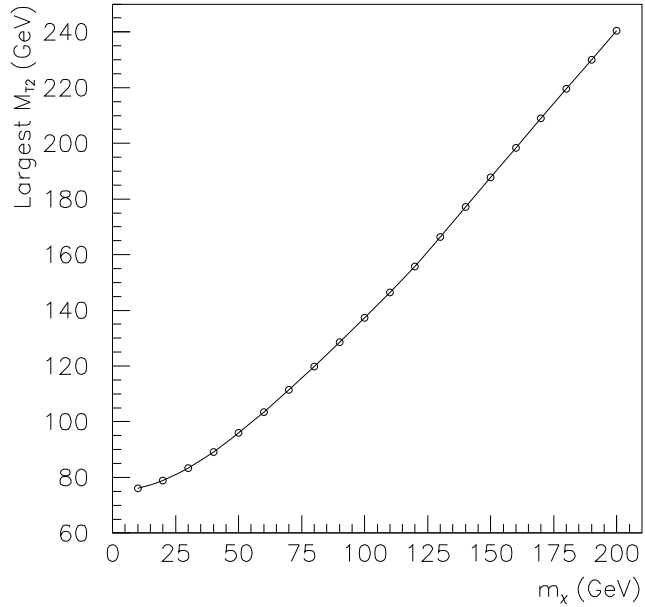,height=8cm}
\end{center}
\caption{Values of $m_{\tilde{l}}$ that would be obtained from that
	largest $M_{T2}$ value observed, where differing values of
	$m_{\tilde{\chi}}$ are used in the calculation of $M_{T2}$,
	for the 1105 events shown in figure \ref{fig:mt2}. All
	events have been generated with $m_{\tilde{l}}=157.1\mbox{ GeV}$ and 
	$m_{\tilde{\chi}}=121.5\mbox{ GeV}$. 
  \label{fig:mlvsmchi}}
\end{figure}

In figure \ref{fig:mt2} we show the $M_{T2}$ distribution for this
model. Quite clearly the maximum $M_{T2}$ value allowed corresponds
well to the mass of the selectron, with the $M_{T2}$ distribution
tending smoothly to zero at that point. While such an edge is not as
easy to measure as a vertical drop, we still expect that such an edge
would be detectable experimentally, and hence a viable means to
measure the selectron mass. The lower bound 
on $M_{T2}$ is $m_{\tilde{\chi}}$, however this does not give a means
to measure the LSP mass, the lower bound on $M_{T2}$ follows
directly from equation (\ref{eq:mtd}), and this is a parameter that goes
into the calculation of $M_{T2}$. Indeed if the value of
$m_{\tilde{\chi}}$ is unknown in addition to $m_{\tilde{l}}$ one can
only obtain a relationship between the two. In figure
\ref{fig:mlvsmchi} we show the value of $m_{\tilde{l}}$ that would
have been obtained from the same events shown in figure \ref{fig:mt2},
using differing input values for $m_{\tilde{\chi}}$ in the
calculation of $M_{T2}$. One can see that slepton mass that one would
extract from the events behaves approximatly as,
\begin{equation}
m_{\tilde{l}} \simeq m_{\tilde{\chi}} + \mbox {constant} ,
\end{equation}
and so any uncertainty in the LSP mass will
be directly reflected as an error in the extracted selectron mass. It should
also be noted that the edge of the $M_{T2}$ distribution becomes hard
to fit as the input $\tilde{\chi}$ mass deviates from the physical
$\tilde{\chi}$ mass as the $M_{T2}$ distributions develops a tail at
the largest $M_{T2}$ values. 

To conclude, in this paper we have introduced a new variable for measuring
masses of particles produced at hadron colliders, where the longitudinal
momentum of the hard scattering is typically unmeasured.  It may be used when
particles are pair produced, with each decaying to one particle that is
directly observed and one particle that is not directly observed. This
variable is analogous to the transverse mass variable, $m_T$, commonly used for
measuring the $W$ mass in its decay to $l\nu$ at $p\bar p$ colliders, except
that it works where the particle being measured is pair produced and where
the unseen particle is massive. We expect that the masses extracted using our
variable, like those from the transverse mass, will be largely independent of
the physics processes which produce the particles, and hence give a viable means
of extracting masses in a model independent way. As an illustration of this
variable in action we consider measuring the mass of selectrons in SUSY at
the LHC. The results look promising however as we have not considered the
effects of any background processes or experimental mis-measurement
errors further study is required\cite{Chris}.

\section*{Acknowledgements}
We would like to thank Andy Parker for helpful conversations.  CGL wishes to
thank his funding body, the Partice Physics and Astronomy Research Council
(PPARC), for financial support.

\vfil\eject

\end{document}